\documentclass[conference,a4paper]{APSIPA2019}
\usepackage{subfigure}
\usepackage{multirow}
\usepackage{graphicx}
\usepackage{amsmath}
\usepackage{amssymb}
\usepackage{amsxtra}
\usepackage{threeparttable}
\usepackage{lipsum}
\usepackage{url}
\usepackage{cleveref}
\usepackage{listings}

\newcommand{\SecLab}[1]{\label{sec:#1}}
\newcommand{\SecRef}[1]{Sec.~\ref{sec:#1}}
\newcommand{\FigLab}[1]{\label{fig:#1}}
\newcommand{\FigRef}[1]{Fig.~\ref{fig:#1}}
\newcommand{\TblLab}[1]{\label{tbl:#1}}
\newcommand{\TblRef}[1]{Table \ref{tbl:#1}}

\begin{document}

\title{Towards Generation of Visual Attention Map \\for Source Code}
\author{
    \authorblockN{
        Takeshi D.~ITOH\authorrefmark{1},
        Takatomi KUBO\authorrefmark{1},
        Kiyoka IKEDA\authorrefmark{1}\authorrefmark{2},
        Yuki MARUNO\authorrefmark{2},
        Yoshiharu IKUTANI\authorrefmark{1},
        \\Hideaki HATA\authorrefmark{1},
        Kenichi MATSUMOTO\authorrefmark{1} and
        Kazushi IKEDA\authorrefmark{1}
    }
    \authorblockA{
        \authorrefmark{1}
        Division of Information Science, Graduate School of Science and Technology,\\
        Nara Institute of Science and Technology, Ikoma, Japan\\
        E-mail: takatomi-k@is.naist.jp
    }
    \authorblockA{
        \authorrefmark{2}
        Department of Contemporary Society, Faculty of Contemporary Society, Kyoto Women's University, Kyoto, Japan
    }
}

\maketitle
\thispagestyle{empty}

\begin{abstract}
Program comprehension is a dominant process in software development and maintenance.
Experts are considered to comprehend the source code efficiently by directing their gaze, or attention, to important components in it.
However, reflecting the importance of components is still a remaining issue in gaze behavior analysis for source code comprehension.
Here we show a conceptual framework to compare the quantified importance of source code components with the gaze behavior of programmers.
We use ``attention'' in attention models (e.g., code2vec) as the importance indices for source code components and evaluate programmers' gaze locations based on the quantified importance.
In this report, we introduce the idea of our gaze behavior analysis using the attention map, and the results of a preliminary experiment.

\end{abstract}

\section{Introduction\SecLab{introduction}}
Program comprehension is a dominant process in software development and maintenance.
Programmers spent 50 \% to 60 \% of their time on program comprehension in a large-scale field study~\cite{Xia18}.
The study has also shown that the senior programmers spent less time on program comprehension.
In other words, the time for program comprehension can be reduced through the appropriate experience or education.
Such efficient program comprehension might lead to productivity enhancement of the software development process.

A clue to improve the efficiency of program comprehension might be the gaze behavior of experts.
We consider that expert programmers can comprehend source code efficiently by directing their gaze, or attention, to important components in it.
In previous studies, researchers conducted gaze measurement experiments with programmers to identify the attended targets.
Uwano et al.~\cite{uwano2006analyzing} analyzed individual performance in reviewing source code of computer programs with gaze data.
Their result showed that the subjects with high performance were likely to first read the whole lines of the source code from the top to the bottom briefly, and then to concentrate their gaze to some particular areas.
Crosby et al.~\cite{crosby2002roles} conducted gaze experiments to examine how programmers from different experience levels understand source code.
Their results showed that experienced programmers were likely to focus their gaze on complex statements.
However, reflecting the importance of components is still a remaining issue in gaze behavior analysis for source code comprehension.

Apart from program comprehension, visual attention has been modeled to clarify its underlying mechanism.
One of the representative studies of visual attention modeling is the saliency map as an indicator of stimulus-driven visual selection~\cite{saliency,2000saliency}.
The saliency map has been proposed by mimicking the neural mechanism of the early visual system of humans.
In the saliency map theory, visual attention is assumed to be guided by high contrast locations of three elementary features: color, intensity, and orientation.
Koide et al. \cite{koide2015art} investigated the relationship between art-related expertise and the saliency map.
They recorded the gaze behavior of artists and novices during the free viewing of various abstract paintings, and evaluated the consistency between the gaze distribution and the saliency map.
The gaze distributions of artists were less consistent with the saliency map than novices.
This discrepancy between the experts' gaze behavior and the saliency map, which is a bottom-up attention model, could be explained by the existence of top-down, goal-oriented attention mechanism which can be modified by experience or education.

This study aims to develop a visual attention map that can identify important components of source code in the top-down, goal-oriented manner.
To do this, we used ``\textit{code2vec}'', a neural network model for classification of source code with an attention mechanism, to identify important components in source code.
Also, we conducted preliminary gaze experiments and a comparison analysis between the gaze behavior of a human subject and the visual attention map generated with our proposed method.
Here, we assume the consistency between the attention map and gaze distribution could be the support for the feasibility of this method to identify important factors.

The rest of this paper is organized as follows:
\SecRef{attention_map} reviews the mechanism of code2vec program summarization model and introduces our idea on how the attention map on top of code2vec's attention model contributes to analyzing programmers' gaze behavior.
\SecRef{experiments} explains the procedure to generate the attention map and gaze experiment design using human subjects.
\SecRef{results} summarizes the results of preliminary experiments.
Finally, \SecRef{discussion} concludes this proposal with some discussion.

\section{Attention Map for Source Code\SecLab{attention_map}}
\begin{figure*}[t]
    \centering
    \includegraphics[width=\textwidth]{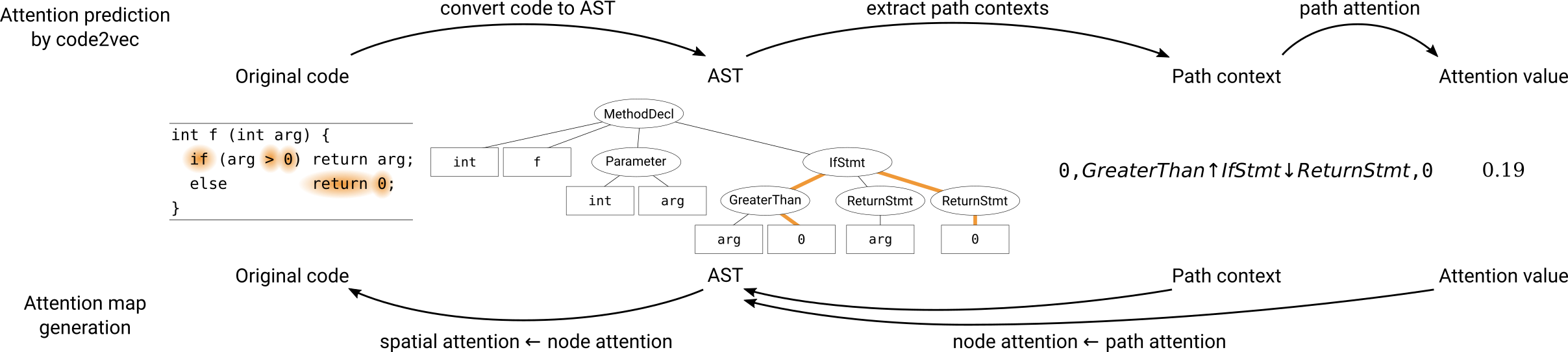}
    \caption{
        Attention estimation by code2vec (top) and proposed attention map generation procedure (bottom).
        The set of orange edges in AST shows an example of a path, and the path attention value (0.19) is assigned to the tokens in the path (orange shadows in the original code).
        The attention values for each token are summed up for all paths by repeating this procedure.
        \FigLab{generation}
    }
\end{figure*}

Code2vec~\cite{alon2019code2vec,code2vec} is a machine learning model to learn a vector embedding of source code, called the ``code vector''.
Code2vec has an attention mechanism to recognize important components in source code for accurate name discrimination.
The model has shown good performance in discriminating function names which concisely represent their functionalities,
and the authors showed the attention mechanism is necessary for achieving good performance.

The upper half of \FigRef{generation} illustrates how code2vec estimates the code vectors and how its attention mechanism defines the importance of source code components.
First, the input source code is converted into an abstract syntax tree (AST), which is a tree data structure representing normalized syntactic information.
Then, code2vec extracts ``path contexts'' from the AST.
A path context consists of three elements: two terminal nodes (leaves) in the AST and the route connecting those terminal nodes.
Code2vec extracts up to 200 path contexts estimates a path context vector for each,
and finally the code vector is computed as a weighted sum of these path context vector (this step is not depicted in \FigRef{generation}).
The ``attention'' of code2vec defines this weight for the path contexts.
A higher attention value means the correspondent path context contains important information for discriminating function name,
and hence the path context vector for such path context greatly affects the final code vector.

In the present study, we assume the expertise of a programmer could be represented using the consistency between the code2vec's attention and a programmer's gaze focus.
Based on this assumption, we evaluate how much a programmer focuses on the source code components that are estimated as important by code2vec's attention mechanism.
However, code2vec's attention is difficult to directly compare with subjects' spatial gaze distribution
since the attention model estimates the importance of path contexts appearing on AST.
To fill the gap, we propose a method to generate spatial ``attention map'' on source code, using the attention value estimated by code2vec.
The detailed procedure of attention map generation is explained in \SecRef{generation}.

\section{Preliminary Experiments\SecLab{experiments}}
\subsection{Acquisition of source code \SecLab{data_acquisition}}

To test the feasibility of visual attention map generation, we conducted a preliminary experiment using a set of code snippets implementing fundamental algorithms.
Based on two popular textbooks about computer algorithms~\cite{Cormen2009Introduction, Sedgewick2011ALG}, we first selected eleven fundamental algorithms: binary search, linear search, bubble sort, selection sort, insertion sort, greatest common divisor, power, primality testing, run-length encoding, string sort, and substring search.
We then collected 1251 Java code snippets implementing the selected algorithms from an open codeset provided by AIZU ONLINE JUDGE~\cite{aizuonlinejudge}.
In the present study, we used a set of 72 code snippets with minimum deviations of superficial characteristics i.e. lines of code (LOC) and characters per line (CPL).
To further mitigate non-semantic visual variations, the indentation styles of all code snippets were normalized by replacing a tab-space with two white-spaces.
For keeping algorithmic diversity, the selected codeset included six snippets for each algorithm but twelve snippets for linear search.
The codeset allowed us to examine the feasibility of our proposed method based on a variety of fundamental algorithms.

\subsection{Attention Map Generation\SecLab{generation}}
To quantify the ``importance'' of each component in the given source code, we used the attention model of code2vec~\cite{alon2019code2vec,code2vec}.
Since code2vec computes the attention for each path context in the given AST of source code,
we reconstructed a spatial attention map based on the attention value of each path context.
The bottom half of \FigRef{generation} illustrates the attention map generation procedure.
In short, we first computed attention values of the nodes appearing in the given AST, then generated a spatial map over a source code image using those node attention values.

First, each path context was decomposed into a list of nodes in the AST.
Then, the attention value for each path context was added to the attention value of each node in this list.
Repeating this procedure for each path context, the attention distribution over the set of AST nodes was obtained.
Next, we converted this node attention distribution into a spatial attention map as a mixture of Gaussian functions.
Some of the nodes had their correspondent tokens in the source code, like \texttt{if} for \textit{IfStmt}, and \texttt{>} symbol for \textit{GreaterThan}.
For those correspondent tokens, a two-dimensional Gaussian function was allocated to each token such that
\begin{enumerate}
    \item its center was located at the coordinate of the center of the token in the stimulus image,
    \item its maximum height was equal to the attention value defined in the attention distribution over AST nodes, and
    \item its variance corresponded to the spatial size of the token in an image.
\end{enumerate}
The spatial map was obtained as the summation of these Gaussian functions for all tokens.

\subsection{Gaze Experiment}
Given the spatial attention map introduced above, the consistency between the map and the programmer's gaze distribution was quantified.
We recorded a programmer's gaze distribution using Tobii Pro TX300~\cite{tobiiprotx300} (Tobii Technology, Sweden),
while presenting the source code image as visual stimuli (see \SecRef{data_acquisition}).
The device has a 23-inch display of full HD resolution (1920 px in width and 1080 px in height).
We recorded the subject's gaze points with the sampling rate of 120 Hz.
The experimental procedure was controlled by PsychoPy~\cite{Peirce2019psychopy2,psychopy}.

After the experiment, the stimulus images were clipped into squares 840 px on a side prior to the further analysis to avoid excessively high consistency due to the inclusion of the blank area in the images.
Outliers in the recorded gaze data which exceed these square boundaries were removed.
The proportion of removed data points against the whole data was less than 0.1 \%.

\subsection{Evaluation}
To quantify the consistency, we adopted an evaluation method proposed in~\cite{koide2015art}.
The receiver operating characteristics (ROC) curve was calculated by defining the ground truth as the subject's gaze distribution and 
the estimation as the binarized attention map computed with code2vec.
Having defined the threshold of attention, the code2vec attention map $C$ was binarized into $C^\mathrm{bin}$:
\begin{equation}
    C^\mathrm{bin}_{x, y} = \begin{cases}
        1 & \mathrm{if}\ C_{x, y} > \mathrm{threshold} \\
        0 & \mathrm{otherwise},
    \end{cases}
\end{equation}
where $C_{x,y}$ and $C^\mathrm{bin}_{x,y}$ represent the attention value for pixel $(x, y)$ in $C$ and $C^\mathrm{bin}$, respectively.
With this binarized attention map, the true positive rate (TPR) and the false positive rate (FPR) given a gaze distribution is calculated as follows.
\begin{align}
    \mathrm{TPR} &= \frac{\sum_{x,y}G^+ \circ C^\mathrm{bin}}{\sum_{x,y}G^+},\\
    \mathrm{FPR} &= \frac{\sum_{x,y}G^- \circ C^\mathrm{bin}}{\sum_{x,y}G^-},
\end{align}
where $G^+$ is the gaze distribution which counts the gaze point per pixel, while $G^-$ is a binary negation of $G^+$, s.t. ${G^-_{x,y} = 1 \Leftrightarrow G^+_{x,y} = 0}$ and $G^-_{x,y} = 0 \Leftrightarrow G^+_{x,y} > 0$.
Also, $\circ$ denotes the Hadamard product (pixel-wise product) of two maps or distributions.
ROC curve was obtained by computing these TPR and FPR with varying the threshold and plotting those values.

After calculating the ROC curve, the area under the curve (AUC) was obtained.
The AUC quantifies the consistency between the subject's gaze distribution and the attention map.
Higher AUC value indicates that the subject strongly focused their gaze on important components in source code,
and thus it is assumed to represent their expertise in reading source code.

\section{Results\SecLab{results}}
\begin{table}[t]
\lstset{
  basicstyle={\ttfamily},
  identifierstyle={\small},
  commentstyle={\smallitshape},
  keywordstyle={\small\bfseries},
  ndkeywordstyle={\small},
  stringstyle={\small\ttfamily},
  frame={tb},
  breaklines=true,
  columns=[l]{fullflexible},
  numbers=left,
  xrightmargin=0em,
  xleftmargin=1.5em,
  numberstyle={\scriptsize},
  stepnumber=1,
  numbersep=1em,
  lineskip=-0.3ex,
  captionpos=b
}
\begin{lstlisting}[caption=A substring search algorithm written in Java,label=javasrc]
public class Main {
  public static void main(String[] args) {
    Scanner in = new Scanner(System.in);
    String word = in.next();
    int count = 0;

    word = word.toLowerCase();

    while (true) {
      String str = in.next();
      if (str.equals("END_OF_TEXT")) {
        break;
      }
      str = str.toLowerCase();
      if (str.equals(word)) {
        count++;
      }
    }
    System.out.println(count);
  }
}
\end{lstlisting}
    \centering
    \caption{Top 5 tokens with strong attention\TblLab{attention_ranking}}
    \begin{tabular}{cccc}
        \# & Line number & Token & Attention value \\ \hline
        1 & 11 & \texttt{if} & 1.33\\
        2 & 11 & \texttt{"END\_OF\_TEXT"} & 1.31 \\
        3 & 9 & \texttt{while} & 1.17 \\
        4 & 2 & \texttt{args} & 0.89 \\
        5 & 7 & \texttt{word}\footnotemark & 0.48
    \end{tabular}
\end{table}
\footnotetext{The one at the left-hand side of the equal symbol.}

\begin{figure*}[htb]
    \vspace{-2mm} 
    \centering
    \subfigure[Spatial attention map]{
        \includegraphics[height=49mm]{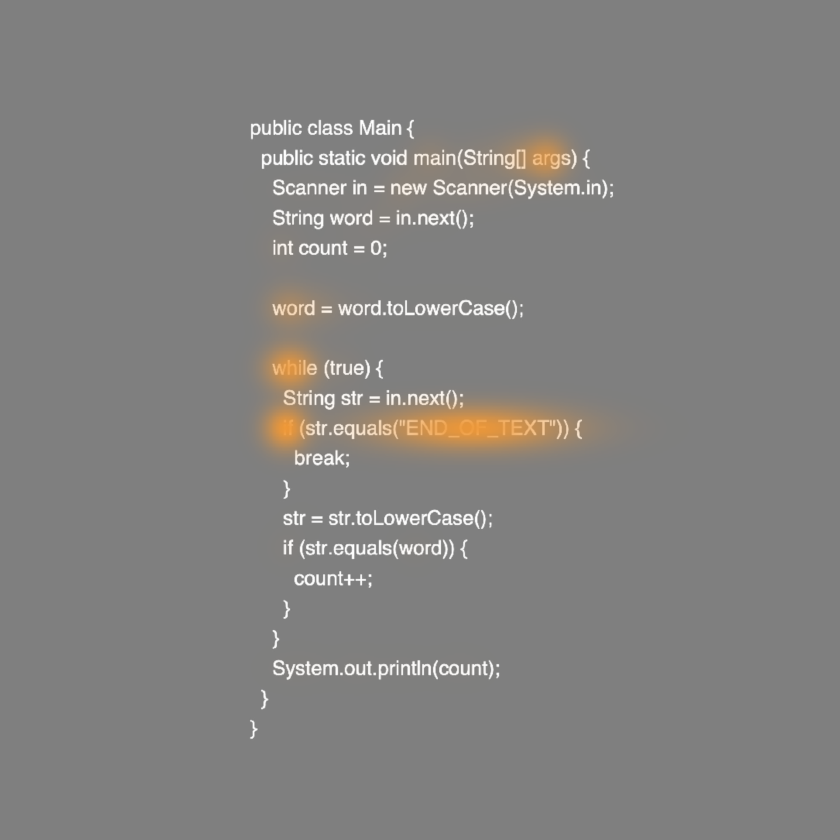}
        \FigLab{attention_map}
    }
    \subfigure[Raw gaze distribution]{
        \includegraphics[height=49mm]{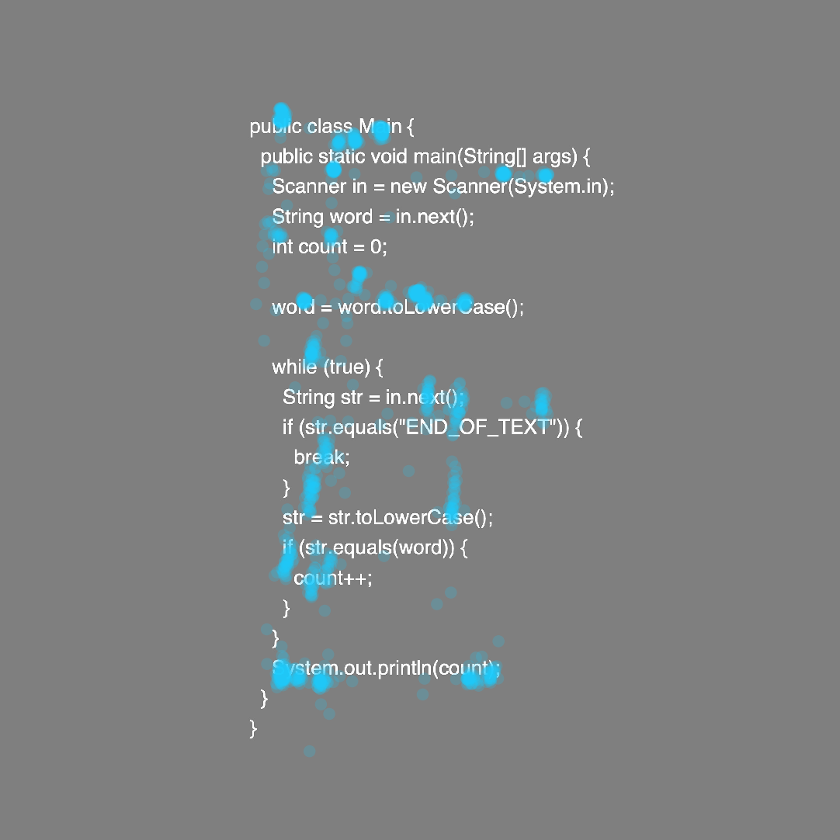}
        \FigLab{raw_gaze}
    }
    \subfigure[ROC curve (AUC=0.85)]{
        \includegraphics[height=49mm]{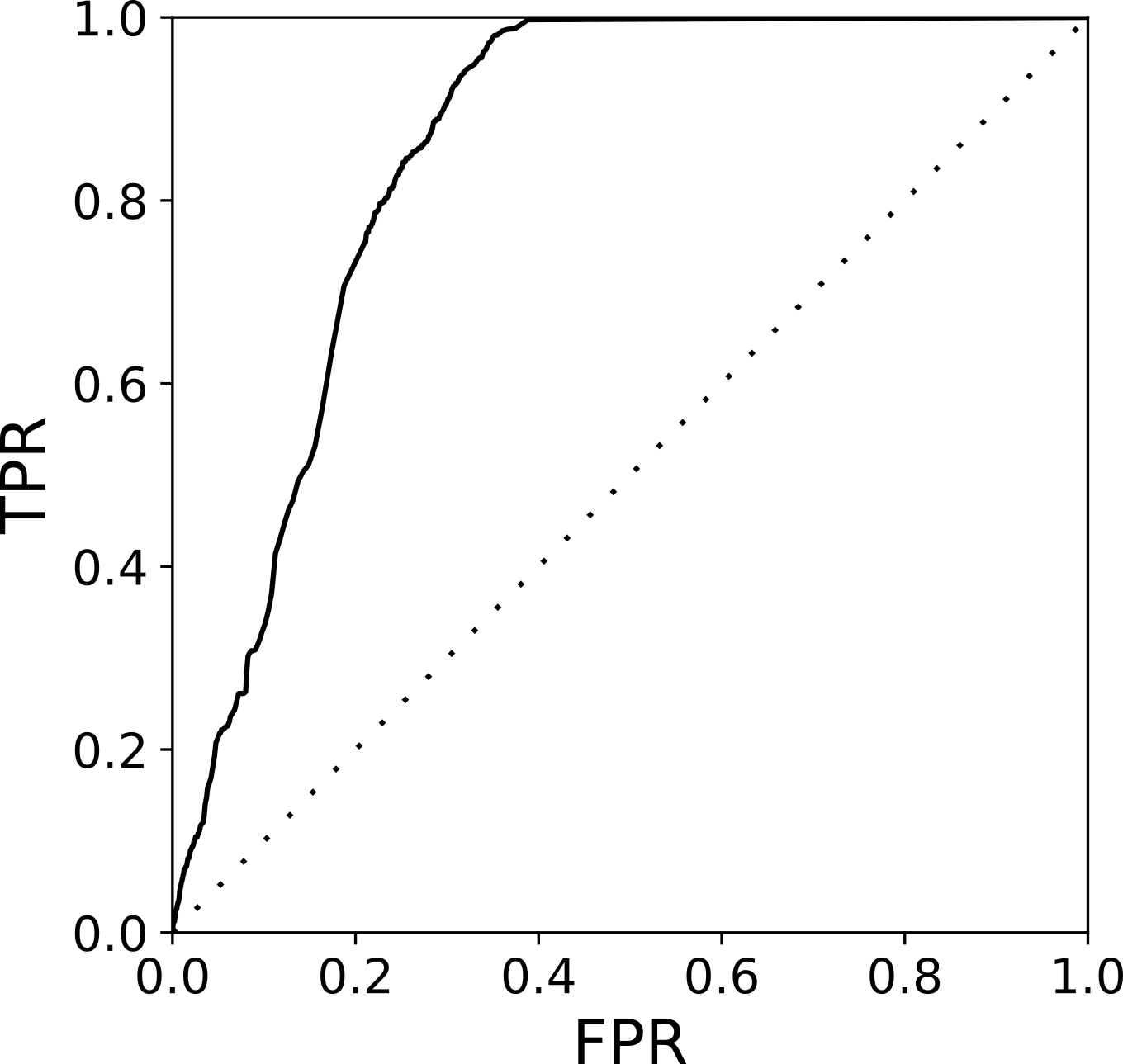}
        \FigLab{roc}
    }
    \caption{An example of result of a gaze experiment on a novice programmer\FigLab{result}}
\end{figure*}

As a preliminary experiment, we quantified the attention maps for the target source code (see \SecRef{data_acquisition}),
and evaluated the consistency of those maps against gaze distributions recorded from a human subject.
In the rest of this section, we describe a representative result obtained using a substring search algorithm shown in Listing \ref{javasrc}.

\FigRef{attention_map} shows the estimated spatial attention map for the source code.
The attention map was sparse, and there were only a few tokens with high attention value (deep orange color).
\TblRef{attention_ranking} lists the top 5 of those highly attended tokens.
Given this source code implements a substring search algorithm that counts the number of words until the word ``END\_OF\_TEXT'' appears,
these highly attended tokens, especially the tokens \texttt{if}, \texttt{while}, and \texttt{"END\_OF\_TEXT"}, match our intuitive evaluation of token importance.

We recorded the gaze behavior of a research student in the information science division who had a little experience in Java programming (i.e. not an expert programmer).
\FigRef{raw_gaze} shows the raw recorded gaze data.
Each blue dot represents a gaze point.
The subject scanned the entire code region without noticeable focuses.

Having the gaze distribution recorded, we evaluated the coincidence between this distribution and the attention map.
The calculated ROC curve is shown in \FigRef{roc}.
The AUC of this ROC curve was 0.85, and this was regarded as moderate consistency.

\section{Discussions\SecLab{discussion}}
We proposed the attention map for source code using code2vec and evaluated the consistency between the attention map and gaze distribution recorded with a subject.
The AUC of 0.85 can be regarded as moderate consistency.
This result supports the feasibility of our proposed method as the attention model to some extent.
The proposed method enables us to evaluate the behavior related to higher cognitive function like program comprehension.
This approach might be applicable to other higher cognitive functions.
On the other hand, the generated visual attention map showed sparser distribution than the gaze distribution.
Since the subject in this experiment was not an expert programmer, gaze distribution was not so concentrated.
Expert programmers may have sparser gaze distributions since such fewer gaze points should lead to a reduction of the time for program comprehension.
In our future experiment, we will recruit multiple expert programmers and novices to show the validity of this attention map.

Other attention models, like code2seq~\cite{alon2018codeseq}, can be alternatives to generate different types of attention maps.
For example, the seq2seq type attention model like code2seq can model the dynamic transition of attention although this study considered static attention~\cite{ikutani2019toward}.
By giving attended words as the output of a decoder, it may be possible to model actual dynamic attention.
Note that introducing attended words to attention in a model is different from the current study
since this study merely evaluated the correspondence of gaze distribution and attention weights of code2vec.
In the future, we will develop such models in parallel with the experimental study mentioned above.

\section*{Acknowledgments}
We would like to thank Dr.~Nishanth Koganti for his valuable advice and comments.
This work was supported by JSPS KAKENHI Grant Numbers JP19J20669, JP18J22957, JP18K18108, JP16H06569, and JP16H05857.

\bibliographystyle{unsrt}

\begin{thebibliography}{10}

\bibitem{Xia18}
Xia Xin, Bao Lingfeng, Lo~David, Xing Zhenchang, E.~Hassan Ahmed, and
  Li~Shanping.
\newblock Measuring program comprehension: A large-scale field study with
  professionals.
\newblock {\em IEEE Transactions on Software Engineering}, 44(10):951--976, Oct
  2018.

\bibitem{uwano2006analyzing}
Hidetake Uwano, Masahide Nakamura, Akito Monden, and Ken-ichi Matsumoto.
\newblock Analyzing individual performance of source code review using
  reviewers' eye movement.
\newblock In {\em Eye Tracking Research and Applications Symposium (ETRA)},
  pages 133--140, 2006.

\bibitem{crosby2002roles}
Martha~E Crosby, Jean Scholtz, and Susan Wiedenbeck.
\newblock The roles beacons play in comprehension for novice and expert
  programmers.
\newblock In {\em 14th Workshop of the Psychology of Programming Interest
  Group}, pages 58--73, 2002.

\bibitem{saliency}
Laurent Itti, Christof Koch, and Ernst Niebur.
\newblock A model of saliency-based visual attention for rapid scene analysis.
\newblock {\em IEEE Transactions on Pattern Analysis and Machine Intelligence},
  20(11):1254--1259, Nov 1998.

\bibitem{2000saliency}
Laurent Itti and Christof Koch.
\newblock A saliency-based search mechanism for overt and covert shifts of
  visual attention.
\newblock {\em Vision research}, 40(10-12):1489--1506, 2000.

\bibitem{koide2015art}
Naoko Koide, Takatomi Kubo, Satoshi Nishida, Tomohiro Shibata, and Kazushi
  Ikeda.
\newblock Art expertise reduces influence of visual salience on fixation in
  viewing abstract-paintings.
\newblock {\em PLOS ONE}, 10(2):1--14, 02 2015.

\bibitem{alon2019code2vec}
Uri Alon, Meital Zilberstein, Omer Levy, and Eran Yahav.
\newblock code2vec: Learning distributed representations of code.
\newblock {\em Proceedings of the ACM on Programming Languages}, 3(POPL):40,
  2019.

\bibitem{code2vec}
code2vec.
\newblock \url{https://code2vec.org}.

\bibitem{Cormen2009Introduction}
Thomas~H. Cormen, Charles~E. Leiserson, Ronald~L. Rivest, and Clifford Stein.
\newblock {\em Introduction to Algorithms, Third Edition}.
\newblock The MIT Press, 3rd edition, 2009.

\bibitem{Sedgewick2011ALG}
Robert Sedgewick and Kevin Wayne.
\newblock {\em Algorithms}.
\newblock Addison-Wesley Professional, 4th edition, 2011.

\bibitem{aizuonlinejudge}
Aizu online judge.
\newblock \url{http://judge.u-aizu.ac.jp/onlinejudge/}.

\bibitem{tobiiprotx300}
Tobii pro tx300 screen-based eye tracker.
\newblock \url{https://www.tobiipro.com/product-listing/tobii-pro-tx300/}.

\bibitem{Peirce2019psychopy2}
Jonathan Peirce, Jeremy~R. Gray, Sol Simpson, Michael MacAskill, Richard
  H{\"o}chenberger, Hiroyuki Sogo, Erik Kastman, and Jonas~Kristoffer
  Lindel{\o}v.
\newblock Psychopy2: Experiments in behavior made easy.
\newblock {\em Behavior Research Methods}, 51(1):195--203, Feb 2019.

\bibitem{psychopy}
Psychopy.
\newblock \url{https://www.psychopy.org}.

\bibitem{alon2018codeseq}
Uri Alon, Omer Levy, and Eran Yahav.
\newblock code2seq: Generating sequences from structured representations of
  code.
\newblock In {\em International Conference on Learning Representations}, 2019.

\bibitem{ikutani2019toward}
Yoshiharu Ikutani, Nishanth Koganti, Hideaki Hata, Takatomi Kubo, and Kenichi
  Matsumoto.
\newblock Toward imitating visual attention of experts in software development
  tasks.
\newblock {\em arXiv preprint arXiv:1903.06320}, 2019.

\end{thebibliography}

\end{document}